\documentclass[12pt,a4paper]{article}
\usepackage{graphicx}
\usepackage{times}
\title{\bf Binaries are the best single stars}

\author{S.E. de Mink$^{1,2}$, N. Langer$^{1,2}$ and R.G. Izzard$^1$\\
\vspace{1cm}\\
\normalsize $^1$ Argelander Institute f\"ur Astronomie der Universit\"at Bonn, Germany\\ 
\normalsize $^2$ Astronomical Institute Utrecht, The Netherlands }
\date{\mbox{}}
\begin{document}
\maketitle
\pagestyle{empty}
%
%
\def\bull{\vrule height .9ex width .8ex depth -.1ex}
\makeatletter
\def\ps@plain{\let\@mkboth\gobbletwo
\def\@oddhead{}\def\@oddfoot{\hfil\tiny\bull\quad
``The multi-wavelength view of hot, massive stars''; 39$^{\rm th}$ Li\`ege Int.\ Astroph.\ Coll., 12-16 July 2010 \quad\bull}%
\def\@evenhead{}\let\@evenfoot\@oddfoot}
\makeatother
%
%
\def\beginrefer{\section*{References}%
\begin{quotation}\mbox{}\par}
\def\refer#1\par{{\setlength{\parindent}{-\leftmargin}\indent#1\par}}
\def\endrefer{\end{quotation}}
%
%
{\noindent\small{\bf Abstract:} 
Stellar models of massive single stars are still plagued by major uncertainties.  Testing and calibrating against observations is essential for their reliability. For this purpose one preferably uses observed stars that have never experienced strong binary interaction, i.e. ``true single stars''.  However, the binary fraction among massive stars is high and identifying ``true single stars'' is not straight forward.   
Binary interaction affects systems in such a way that the initially less massive star becomes, or appears to be, single. For example, mass transfer results in a widening of the orbit and a decrease of the luminosity of the donor star, which makes it very hard to detect. After a merger or disruption of the system by the supernova explosion, no companion will be present.  

%
The only unambiguous identification of ``true single stars'' is possible in detached binaries, which contain two main-sequence stars. For these systems we can exclude the occurrence of mass transfer since their birth. 
A further advantage is that binaries can often provide us with direct measurements of the fundamental stellar parameters. Therefore,  we argue these binaries are worth the effort needed to observe and analyze them.  They may provide the most stringent test cases for single stellar models. 

}
%
%


\section{Introduction}
{``Massive stars appear to love company'.'}   With this sentence Mason et al. (2009) open and summarize their paper describing a comprehensive compilation of spectroscopic data of close binaries and high angular resolution data of wide binaries.   They conclude that more than half of the stars in the Galactic O-star catalogue are spectroscopic binaries.  Using a smaller, but homogeneously analyzed data set, Sana \& Evans (2010) find a spectroscopic binary fraction of  $44\pm5\%$ for nearby clusters that are rich in O-stars. As these authors phrase it: ``to ignore the multiplicity of early-type stars is equivalent to neglecting one of their most defining characteristics'', see also Sana et al. (2008).

Spectroscopic measurements can identify binaries with separations up to a few AU or orbital periods up to a few years.  This is of the order of the maximum separation and orbital period for which binaries are close enough to interact by mass transfer.   In such close binaries the presence of a nearby companion can drastically alter the further evolution, the observable properties and the final fate of both stars (e.g. Kippenhahn \& Weigert 1967, Podsiadlowski, Joss \& Hsu 1992, Pols 1994, Wellstein \& Langer 1999, Eldridge, Izzard \& Tout 2008).

Besides the complexity of the physics of binary interaction, we have to face the fact that stellar models of massive single stars are still plagued by major uncertainties.  Even during one of the simplest evolutionary phases, the main-sequence evolution, their evolution is strongly affected by poorly constrained internal mixing processes and mass loss.   These uncertainties affect all subsequent evolutionary phases and therefore the large role that massive stars play in the enrichment of the interstellar medium, as sources of ionizing radiation and as progenitors of supernovae and gamma-ray bursts. 

Promising opportunities to calibrate and test stellar models come from large, homogeneously analyzed samples. For example, the VLT-flames survey of massive stars (Evans et al. 2005), quantified the metallicity dependence of stellar winds (Mokiem et al. 2007). It also provided the first homogeneously analyzed data set of surface abundance measurements for stars exhibiting a wide range of rotational velocities, which is crucial for testing the effects of rotationally induced mixing processes (Hunter et al. 2009). 

When performing such tests it is important to consider whether the observed properties are indeed the result of single star evolution or whether they might have been caused by interaction with a binary companion.  For example, a rapidly rotating star with peculiar surface abundances can be interpreted as evidence that certain mixing processes operate in rotating single stars (Hunter et al. 2009, Maeder et al. 2009). Alternatively, its properties can be interpreted as a signature of a previous phase of mass accretion from a binary companion (Langer et al. 2008).  
As this example shows,  single stellar physics can in some cases lead to similar observable signatures as expected for binary interaction.   Because of this degeneracy it is important to test the physics of single stellar models using observations of stars that have not experienced strong binary interaction such as mass transfer.   Identifying such stars is not a simple task because of the effects of binary interaction are  somewhat counter-intuitive.  In Section~2 we explain why stars that have experienced strong binary interaction often appear to be single.  In Section~3 we argue that ``true single stars'',  can be found in binaries.

\section{ Not every single star was always single }
Absence of evidence for a companion star does not guarantee that the star has never experienced binary interaction.   In fact, a star that has experienced binary interaction in the past will often be -- or appear to be -- single.  To illustrate this we present preliminary results of computations with a rapid binary evolutionary code (Hurley et al.~2002, Izzard et al.~2006, De Mink et al.\ in prep.).   For the purpose of the study of massive binaries we updated the treatment of mass and angular momentum loss and transfer, by implementing an improved Roche-lobe overflow scheme to determine the mass-transfer rate.  Furthermore, we include a treatment of the effects of rotation on the stellar structure and the mass-loss rate.  We calibrated the code against grids of models computed with the detailed binary evolution codes {\tt STARS} (Eggleton 1971, Pols et al. 1995, De Mink, Pols \& Hilditch 2007) and {\tt BEC} (e.g. Wellstein \& Langer 1999, Petrovic et al. 2005). 

\begin{figure}[t]
\centering
\includegraphics[width=\textwidth]{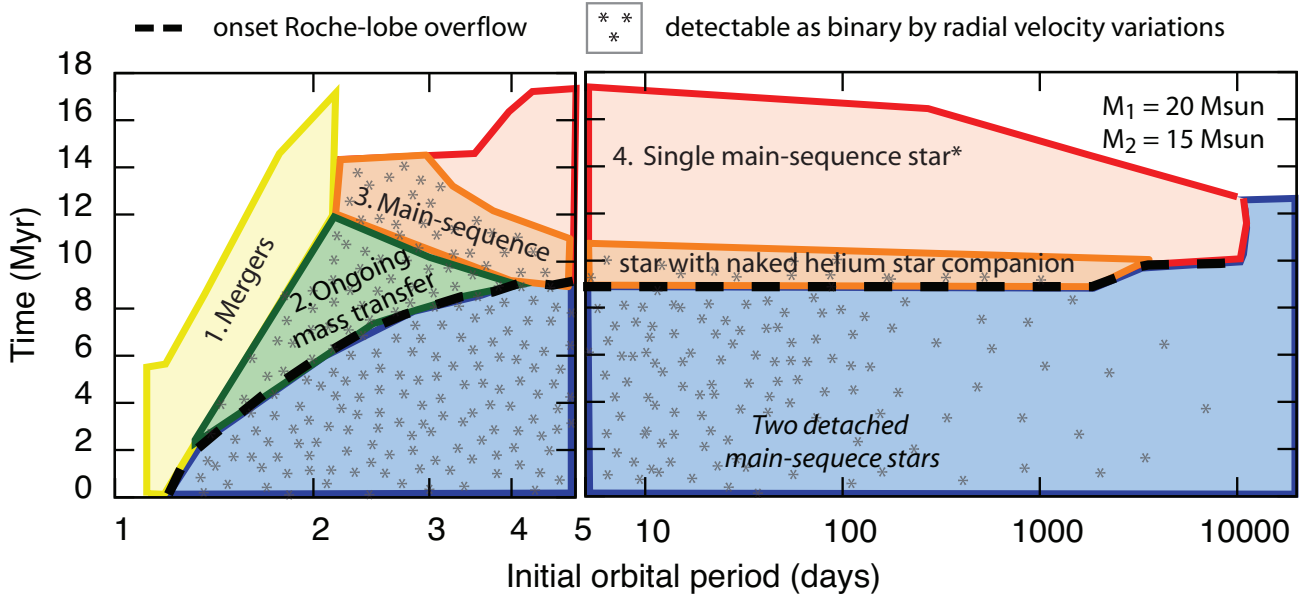}
\caption{Schematic depiction of the evolutionary stages of a 20+15$M_{\odot}$ binary as a function of the initial orbital period ($x$-axis) and time ($y$-axis). In the left panel we zoom in on very close binaries. We only the depict phases in which at least one of the stars is on the main-sequence.  The thick dashed line indicates the onset of mass transfer.  Plusses indicate qualitatively during which phases systems are detectable as spectroscopic binaries.  See text for more explanation. \label{fig:logp-time}}
\end{figure}

In Figure~\ref{fig:logp-time} we depict the evolutionary stages of a massive binary adopting initial masses of 20 and 15 solar masses for the primary and secondary star. We vary the initial orbital period.  The evolution of such systems is representative for massive binaries with a primary mass between roughly 10 and 40 solar masses, with not too extreme mass ratios. The figure is intended as illustration of the main argument, not as a quantitative prediction.  We discuss different phases for which at least one of the stars is a main-sequence star. 

\paragraph{Pre-interaction phase}Initially both stars reside within their Roche lobes and the binary system is in a detached configuration (blue area).  For systems with orbital periods larger than 4 days, this phase lasts approximately as long as the main-sequence lifetime of the primary star.  In tighter systems the initially most massive star fills its Roche lobe before finishing its main-sequence evolution (thick dashed line). 
During this phase the stars evolve similarly to single stars. Their interaction in limited to interaction via stellar winds and tides.  In systems with orbital periods smaller than about 10 days the stars tides will affect their rotation rate. Besides this the physical processes in the star are not expected to differ significantly from those in single stars, assuming that tidally induced mixing processes  and angular momentum transport are not very efficient. We refer the reader to the De Mink et al. (2009) for a study of the consequences of mixing processes in tidally locked binaries.

These systems can in principle be detected as double-lined spectroscopic binaries, depending on the timing of the observations and on the inclination and eccentricity of the orbit (Sana, Gosset \& Evans, 2010).  A simulation of the detection probability of binaries in the VLT flames Tarantula survey shows that nearly all binaries with orbital periods up to a 100 days can be identified  ($>$90\%).  The fraction decreases for wider binaries, but one still expects to recover about half of the binaries with orbital periods between 100 and 1000 days (Evans et al. 2010).

\paragraph {Mergers [1] }
In the tightest binaries mass transfer will lead to a merger of the two stars  (indicated in yellow in Fig.~\ref{fig:logp-time}).  At the moment of contact  both stars have only partially burned the hydrogen in their center.  The merger product will therefore also be a hydrogen burning, main-sequence star.   
Although the very tight binaries that lead to these kind of mergers are not very common, there are reasons to believe that mergers can nevertheless not be ignored.   In old star clusters mergers can be identified as blue stragglers, being significantly brighter and bluer than stars near the main-sequence turn-off.  In young star clusters the turn-off is not well defined and blue stragglers cannot clearly be identified.  Given the high fraction of close binaries among massive stars, mergers may constitute a very significant fraction of the stars near the turn-off in young clusters.    In regions with continuous star formation mergers will be as bright as newly formed massive single stars.  Mergers result from lower mass stars, which are more abundant due to the slope of the initial mass function.  Therefore, the fraction of mergers among stars within a given luminosity range  may still be significant.

Mergers will have no companion, unless the system was originally a triple.  Therefore mergers may be easily mistaken for a primordial single star even though their evolutionary history is completely different. Unfortunately the merger process is poorly understood. In particular, internal mixing, mass and angular momentum loss are uncertain. It is not yet clear whether these mergers demonstrate peculiar observable properties, for example larger rotation rates and enhanced surface nitrogen abundances.  Attributing their properties to single stellar physics might lead to erroneous conclusions.

\paragraph{Semi-detached systems [2]}
Binary systems in which the primary fills its Roche lobe during its main-sequence evolution experience a long-lasting phase of mass transfer (indicated in green in Fig~\ref{fig:logp-time}).  
As these systems are close, tides are efficient in locking the rotation of the both stars to the orbit.  Large amounts of mass can be accreted onto the companion star, while the transferred angular momentum is quickly converted into orbital angular momentum due to tides (e.g. Petrovic et al.~ 2005).  
%
%
%
%
Because of their geometry and short orbital periods, these systems will show eclipses and radial velocity variations in nearly all cases.  They can be  useful to test the physics of mass transfer and contact systems (e.g. De Mink et al. 2007, van Rensbergen et al. 2010).

\paragraph{Post-interaction: a stripped helium star companion [3]}
When the initially most massive star fills its Roche lobe it will continue to lose mass until it has been stripped from its entire hydrogen-rich envelope.  For systems with initial orbital periods larger than 4 days, the mass-transfer phase lasts only a thermal timescale. This is short compared to the nuclear timescale and is not visible in Fig.~\ref{fig:logp-time}. During the mass-transfer phase the orbit widens and the secondary is now the brightest star in the system.  The stripped primary star is very hard to detect, due to its low luminosity, low mass and the wide orbit.  In addition, the secondary is expected to be rotating rapidly as a result of the accreted angular momentum.  Broadening of the spectral lines because of rotation makes it even harder to detect spectral lines of the companion.  Therefore, these objects often appear to be single even though their evolution is severely affected by binary interaction.

The duration of this phase, indicated in orange in Fig.~\ref{fig:logp-time}, is set by the helium burning lifetime of the primary star.   This is in the order of one tenth of its main-sequence lifetime, or longer as stellar winds reduce the mass of the helium star.   Even though the duration of this phase is considerable, to our knowledge no such massive binary has been detected, which may reflect the difficulty to detect such systems.   

\paragraph {Post-interaction: disrupted systems and neutron star companions [4]}
By the time the initially most massive star explodes, the orbit has become fairly wide due to mass transfer and stellar winds.  In most cases, the binary is expected to be disrupted, leaving the secondary behind as a single star.  The secondary star may acquire a moderate spacial velocity, although the formation of runaway stars will be the exception rather than the rule.   This phase, indicated in red in Fig.~\ref{fig:logp-time}, lasts as long as the remaining main-sequence lifetime of the secondary.

\section{Binaries provide our only chance to identify true single stars}

We argued above that many stars are, or appear to be, single after experiencing strong binary interaction.  Such stars are expected to constitute a sizable fraction of all single stars and can not be neglected.  Searching for stars without a companion is therefore not an effective method to identify ``true single stars'', i.e.\ stars that have lived their lives without experiencing strong interacting such as mass transfer.
Similarly, removing stars with evidence for binarity from an observed sample will be counter-productive.  One would preferentially remove the pre-interaction systems, in which stars have lived their lives similarly to single stars, while post-interaction systems are left in the sample.   In other words, excluding detected binaries from a sample increases the relative fraction of stars that have been affected by binary interaction.

Fortunately, binaries provide us with an opportunity to identify stars that have not suffered from strong binary interaction.  Evolutionary models show that binary interaction by Roche-lobe overflow strips the donor star from its hydrogen-rich envelope.  When one of the stars fills its Roche lobe, it does not detach again until the entire envelope has been removed.  Therefore, in a close binary system containing two detached main-sequence stars, we can exclude the occurrence of mass transfer since their birth as hydrogen burning stars.   

Such binaries often provide us with an unique method to measure the fundamental stellar parameters directly.  For this reason they have been used successfully by various authors to test evolutionary models of single stars (e.g. Schr{\"o}der et al. 1995, Pols et al. 1995, Pavlovski \& Southworth 2009, Pavlovski et al 2009).   Therefore we advocate that, in order to increase our understanding of single stars, binaries provide one of our best opportunities.

\section*{Acknowledgements}
We acknowledge the members of the VLT flames consortium of massive stars and the stellar evolutionary groups in Bonn and Utrecht for fruitful discussions. 
%
%



\footnotesize
\beginrefer

\refer De Mink S.~E., Pols O.~R., Hilditch R.~W., 2007, A\&A, 467, 1181

\refer De Mink, S.~E., Cantiello, M., Langer, N., Pols, O.~R., Brott, I., \& Yoon, S.-C.\ 2009,  A\&A, 497, 243 

\refer Eggleton P.~P., 1971, MNRAS, 151, 351 

\refer Eldridge J.~J., Izzard R.~G., Tout C.~A., 2008, MNRAS, 384, 1109 

\refer Evans, C.~J., Smartt, S.~J., Lee, J.-K.,  et al.\ 2005, A\&A, 437, 467 

\refer Evans, C.~J., Bastian, N., Beletsky, Y., et al.\ 2010, IAU Symposium, 266, 35 

\refer Hunter, I., Brott, I., Langer, N., et al.\ 2009, A\&A, 496, 841 

\refer Hurley, J.~R., Tout,  C.~A., \& Pols, O.~R.\ 2002, MNRAS, 329, 897 

\refer Izzard, R.~G., Dray, L.~M., Karakas, A.~I., Lugaro, M., \& Tout, C.~A.\ 2006, A\&A, 460, 565 

\refer Kippenhahn, R., \& Weigert, A.1967, ZAP 65, 251

\refer Langer N., Cantiello M., Yoon S.-C., Hunter I., Brott I., Lennon D., De Mink S.~E., Verheijdt M., 2008, IAU Symposium, 250, 167  

\refer Maeder, A., Meynet, G.,  Ekstr{\"o}m, S.,  \& Georgy, C.\ 2009, Comm. in Asteroseismology, 158, 72 

\refer Mason B.~D., Hartkopf W.~I., Gies D.~R., Henry T.~J., Helsel J.~W., 2009, AJ, 137, 3358 

\refer Mokiem, M.~R., de Koter, A., Vink, J.~S., et al.\ 2007, A\&A, 473, 603 

\refer Pavlovski, K., \& Southworth, J.\ 2009, MNRAS, 394, 1519 

\refer Pavlovski, K.,  Tamajo, E., Koubsk{\'y}, P., Southworth, J., Yang, S.,  \& Kolbas, V.\ 2009, MNRAS, 400, 791 

\refer Petrovic, J., Langer, N., Yoon, S.-C., \& Heger, A.\ 2005, A\&A, 435, 247 

\refer Podsiadlowski P., Joss P.~C., Hsu J.~J.~L., 1992, ApJ, 391, 246 

\refer Pols O.~R., 1994, A\&A, 290, 119 

\refer Pols O.~R., Tout C.~A., Eggleton P.~P., Han Z., 1995, MNRAS, 274, 964

\refer Pols O.~R., Tout C.~A., Schr{\"o}der K.-P., Eggleton P.~P., Manners J., 1997,  MNRAS, 289, 869 

\refer Sana H., Gosset E., Naz{\'e} Y., Rauw G., Linder N., 2008, MNRAS, 386, 447 

\refer Sana H., Gosset E., Evans C.~J., 2009, MNRAS, 400, 1479 

\refer Sana H., Evans C.~J., 2010, arXiv, arXiv:1009.4197 

\refer  Schr{\"o}der K.-P., Pols O.~R., Eggleton P.~P., 1997, MNRAS, 285, 696 

\refer  van Rensbergen, W., De Greve, J.~P., Mennekens, N., Jansen, K., \& De Loore, C.\ 2010, A\&A, 510, A13 

\refer Wellstein S., Langer N., 1999, A\&A, 350, 148 


\endrefer           
\end{document}